# Fabrication and some properties of biaxially aligned $Sr_{0.6}K_{0.4}Fe_2As_2$ superconductors by processing in high magnetic field

Zhaoshun Gao[1], Xianping Zhang[1], Dongliang Wang[1], Yanpeng Qi[1], Lei Wang[1], Junsheng Cheng[1], Qiuliang Wang[1], Yanwei Ma[1]*, S. Awaji[2], K. Watanabe[2]

[1] Key Laboratory of Applied Superconductivity, Institute of Electrical Engineering, Chinese Academy of Sciences, Beijing 100190, China

[2] High Field Laboratory for Superconducting Materials, Institute for Materials Research, Tohoku University, Sendai 980-8577, Japan

**Abstract:**

We fabricated the *c* axis and *ab*-plane biaxially aligned $Sr_{0.6}K_{0.4}Fe_2As_2$ superconductor using a two-step magnetic field procedure. The effect of magnetic fields on the structure and superconducting properties of $Sr_{0.6}K_{0.4}Fe_2As_2$ has been investigated by using X-ray diffraction and magnetic measurements. The degree of orientation of the samples was about 0.39 for the *c* axis and 0.51 for *ab*-plane orientation, as evaluated from the Lotgering factor of X-ray diffraction. This technology might be useful in a variety of potential applications, including preparing iron based superconducting bulks and wires with high critical currents.

* Author to whom correspondence should be addressed; E-mail: ywma@mail.iee.ac.cn

The discovery of superconductivity in the iron pnictides has triggered great interest from both theoretical and application point of view [1-7]. In addition to the high transition temperature, $T_c$, these FeAs based superconductors were reported to have a very high upper critical field, $H_{c2}$, bringing the hope in a wide array of future applications [8–10]. However, soon after the discovery of the iron based superconductors, the issue of weak link was observed in polycrystalline samples, showing that the intergrain and intragrain current densities differ in magnitude by about $10^3$ [10-14]. Recently, bicrystal grain boundary experiments [15, 16] indicated that the weak link behavior of high-angle grain boundary was crucial for affecting the critical current $J_c$ in iron-based superconductors. Those results suggest that the newly discovered FeAs based superconductors are exhibiting weak link grain boundary behavior similar to high-$T_c$ cuprate superconductors. As we know, this limitation is strongly correlated with the misorientation between the grains. In order to avoid the presence of those weak links in the materials, one possibility is to texture the constitutive grains in such a way that the super current may pass through the intergrain barriers without loss of the superconducting wave-function coherence [17].

Processing in an external magnetic field is a well-proven technique to enhance the degree of grain alignment and critical current density for the case of high-$T_c$ oxide superconductors [18-19]. The driving force for grain alignment is provided by the anisotropic paramagnetic susceptibility exhibited by the superconductor grains. When a superconductor grain is placed in a magnetic field, the axis of maximum susceptibility aligns with the magnetic field direction to minimize its energy in the field. It is known that the FeAs layer structure of iron based superconductors is anisotropic [1]. If a magnetic field is applied during the magnetic process, grain alignment is expected. Actually, more recently, grain alignment in REOFeAs(1111) compound of iron based superconductors induced by a magnetic field has been reported in [20, 21], little attention has been paid to the case of $AFe_2As_2$(122) phases processed under the same conditions. The present work deals with the possibility to align $Sr_{0.6}K_{0.4}Fe_2As_2$ under a magnetic field.

Polycrystalline samples of $Sr_{0.6}K_{0.4}Fe_2As_2$ were prepared by a one-step PIT method developed by our group [22]. The samples were ground to a powder in an agate mortar and pestle, mixed with epoxy resin in a Teflon tube of diameter 4 mm with a typical powder to epoxy ratio of 1:5 and then aligned in an 8 T superconducting magnet at room temperature. We have performed the alignment at room temperature using a two-step magnetic field procedure to obtain grain orientation, as illustrated in Fig. 1. In the first step, field was applied perpendicular to the tube axis, which produced a sample with *c* axes randomly distributed in the plane perpendicular to the applied field (Fig. 1a). After about half hour, the tube was rotated 90° about its long axis (Fig. 1b) and the epoxy was allowed to set. As a result, particles reoriented to align with a *c* axis.

The degree of alignment was checked by means of x-ray diffraction analysis. Figure 2 illustrates XRD profiles of the samples using first step direction of the strong magnetic field and the samples prepared without a magnetic field. In the surface perpendicular to the magnetic field in Fig. 2a, the intensities of the (200) and (220) peaks were very large compared to the bottom profile of the sample with a random orientation. On the other hand, in the surface parallel to the magnetic field in Fig. 1b the intensity of the (200) peak was weak. In this case, the *ab*-plane aligned parallel to the direction of the magnetic field. The XRD analysis revealed that the *ab*-plane was the easy magnetization axis for $Sr_{0.6}K_{0.4}Fe_2As_2$, because peak intensities were enhanced associated with the *ab*- plane of the grains such as (200) and (220) peaks.

Fig. 3 shows X-ray diffraction patterns for samples prepared at second step, and the samples prepared without a magnetic field. The analysis was the same as above. An enhanced peak for the *c*-axis of the grain was observed in the 8 T static magnetic field, but small peaks associated with the *ab*-plane were also observed (Fig. 3(a)). In the face perpendicular to the magnetic field, peaks associated with the *ab*-plane of the crystal, such as (100) and (110), were presnt (Fig. 3(b)). By contrast, in the surface parallel to magnetic field, the intensity of the (002) peak was strong. Therefore the *c* axis of the grains aligned perpendicular to magnetic field. The preferred orientation factor F is evaluated by the Lotgering method as follows [23].

$$F=(\rho - \rho_0)/(1 - \rho_0),$$

$\rho_0 =\sum I_0(00l)/ \sum I_0(hkl)$, $\rho =\sum I(00l)/ \sum I(hkl)$ for *c* axis orientation, and $\rho_0 =\sum I_0(hk0)/ \sum I_0(hkl)$, $\rho =\sum I(hk0)/ \sum I(hkl)$ for ab-plane orientation, respectively. Where $I$ and $I_0$ are the intensities of each reflection peak (*hkl*) for the oriented and random samples, respectively. The value of F for the *c* axis orientation was 0.39, and the value of F for ab-plane orientation was 0.51, respectively. This clearly demonstrates that a crystalline orientation of the *c* axis, as well as *ab*-plane, is developed simultaneously in $Sr_{0.6}K_{0.4}Fe_2As_2$ using two-step magnetic field procedure.

The anisotropic temperature dependences of the molar magnetic susceptibilities $\chi_{ab}(T)$ and $\chi_c(T)$ for the aligned $Sr_{0.6}K_{0.4}Fe_2As_2$ powder with applied field along the *ab*-plane and the *c*-axis are shown collectively in fig. 4. For an aligned dispersed microcrystalline sample at a low applied field of 20 Oe, both zero-field–cooled (ZFC) and field-cooled (FC) data revealed a superconducting transition temperature $T_c$ of 35 K, which is identical to the $T_c$ measured from the bulk polycrystalline sample [22]. The anisotropic diamagnetic parameter $\chi_c/\chi_{ab}$ of 1.22 at 5 K was deduced for the aligned microcrystalline. The small anisotropy is consistent with the results reported in Ref [24].

The samples with the dimension of 2 mm×2 mm×1.5 mm were cut from the pellets after treatment in the high magnetic field to measure magnetic hysteresis loops by a vibrating sample magnetometer. The samples were set in two directions in the measurement, with the measuring field $H_m$ parallel to *ab*-plane and *c*-axis respectively. The result obtained at 5 K for the sample is shown in Fig. 5. The hysteresis loop of $H_m$ parallel to *c*-axis is slightly wider than that of $H_m$ parallel to *ab*-plane. Obviously this difference in the hysteresis loops is due to the grain alignment in the magnetic field.

The easy magnetization axis and a minimum value of applied magnetic field required for grain alignment were initially determined. Due to the anisotropic FeAs layer structure, anisotropic Fe magnetic susceptibility $\chi_{ab}(Fe)>\chi_c(Fe)$ is expected to be the dominant factor for the anisotropic magnetic susceptibility of $Sr_{0.6}K_{0.4}Fe_2As_2$[20]. The magnetic susceptibilities along the *ab*-plane and *c*-axis for $SrFe_2As_2$ were reported as an order of $10^{-6}$ emu/cm$^3$ [25]. Assuming that the magnetic susceptibility of $Sr_{0.6}K_{0.4}Fe_2As_2$ is at the same order as that of $SrFe_2As_2$, and then an anisotropic magnetic susceptibility, $\Delta\chi = \chi_{\parallel} - \chi_{\perp}$ where $\chi_{\parallel}$ and $\chi_{\perp}$ are the susceptibilities parallel and perpendicular to the magnetic principal axis, respectively, should be on the order between $10^{-7}$ and $10^{-9}$ for $Sr_{0.6}K_{0.4}Fe_2As_2$. Thus, the orientation energy estimated for a 5 μm spherical and single crystalline particle ranges from $10^{-17}$ to $10^{-19}$ J in an 8 T magnetic field. On the other hand, the kinetic energy due to thermal motion ($kT$) at 300 K is $4.14 \times 10^{-21}$ J, where $k$ is Boltzmann's constant. These values suggest that the orientation energy of 8 T is large enough to rotate the powders.

In conclusion, these experiments confirm that $Sr_{0.6}K_{0.4}Fe_2As_2$ can be well oriented in a magnetic field. Therefore, magnetic texturing growth of iron based superconductors should be envisaged for improving electric percolation paths. A further development of the technique (removable binder and appropriate thermal treatment) might facilitate the production of aligned bulk and wire samples with much improved critical currents.

The authors thank Xiaohang Li, Hui Wang and Yuangzhong Lei for their help and useful discussion. This work is partially supported by Beijing Municipal Natural Science Foundation (No. 2102045), the National Natural Science Foundation of China under Grant No. 51002150 and National '973' Program (Grant No. 2006CB601004).

# Captions

Figure 1  A schematic illustration shows the two-step magnetic field procedure. (a) the first step produces particles with *ab* plane parallel to external field and their *c* axis randomly oriented in a plane, perpendicular to the field. And (b) After 90° rotation, the second step aligns *c* with the sample's long axis.

Figure 2  The x-ray diffraction patterns of the samples prepared in the magnetic field at first step and without a magnetic field.

Figure 3  The x-ray diffraction patterns of the samples prepared in the magnetic field at second step and without a magnetic field.

Figure 4  Anisotropic ZFC and FC susceptibilities $\chi_{ab}(T)$ and $\chi_c(T)$ for aligned $Sr_{0.6}K_{0.4}Fe_2As_2$ powder.

Figure 5  Anisotropic high-field superconducting hysteresis loop $M_cH$ and $M_{ab}H$ at 10 K for the aligned $Sr_{0.6}K_{0.4}Fe_2As_2$ powder.

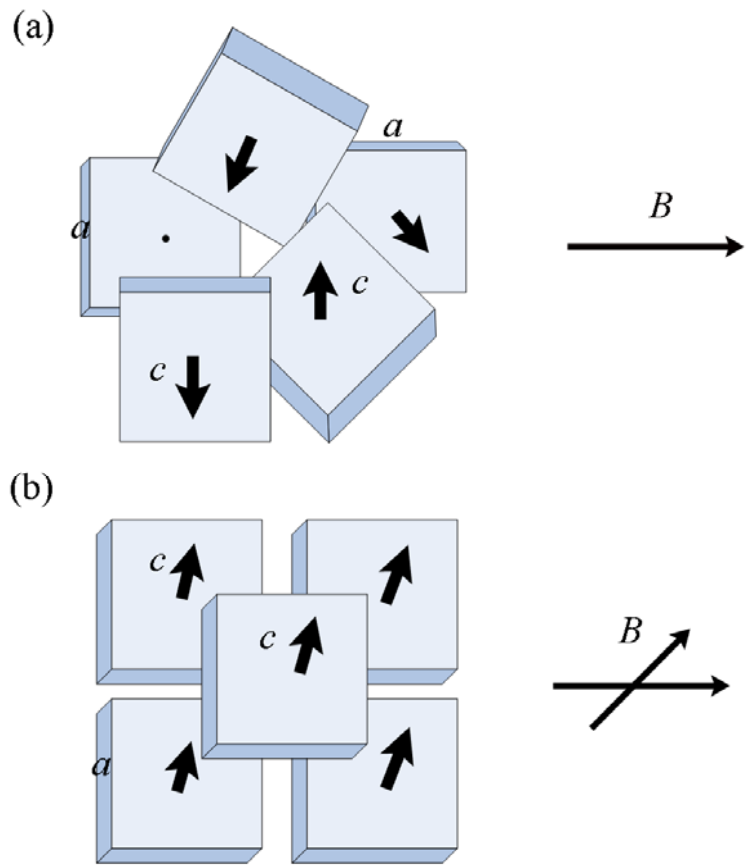

Fig.1 Gao et al.

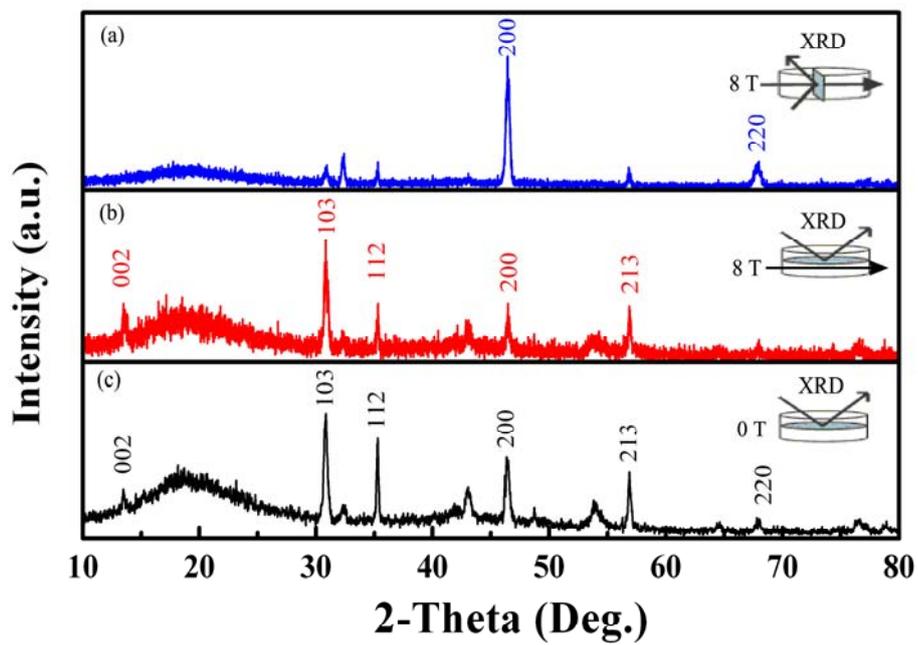

Fig.2 Gao et al.

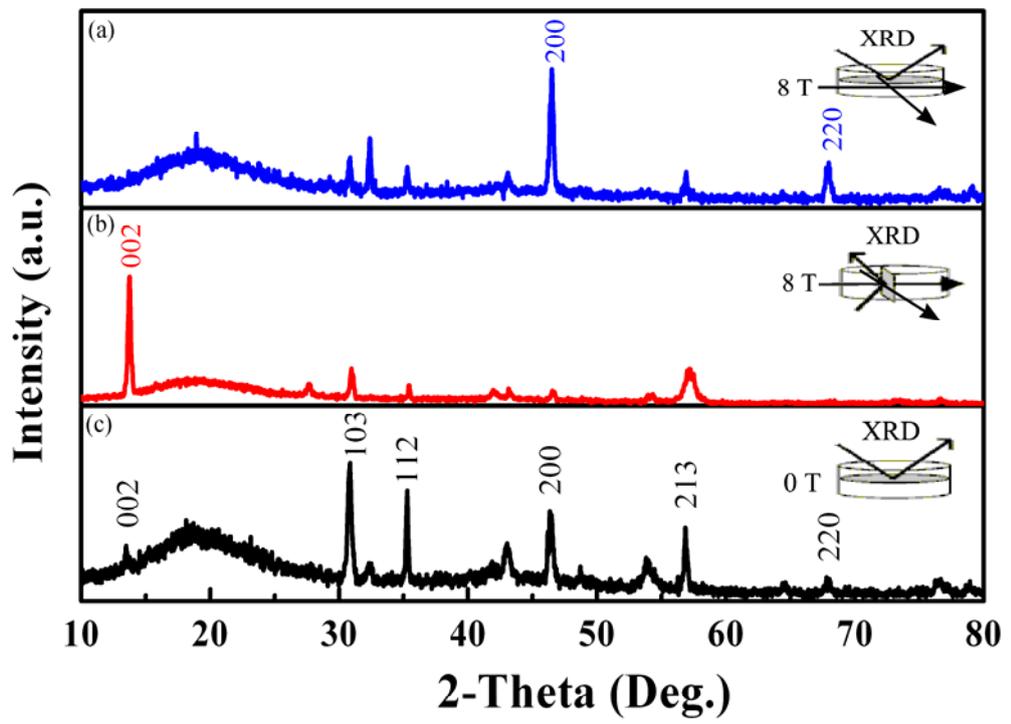

Fig.3 Gao et al.

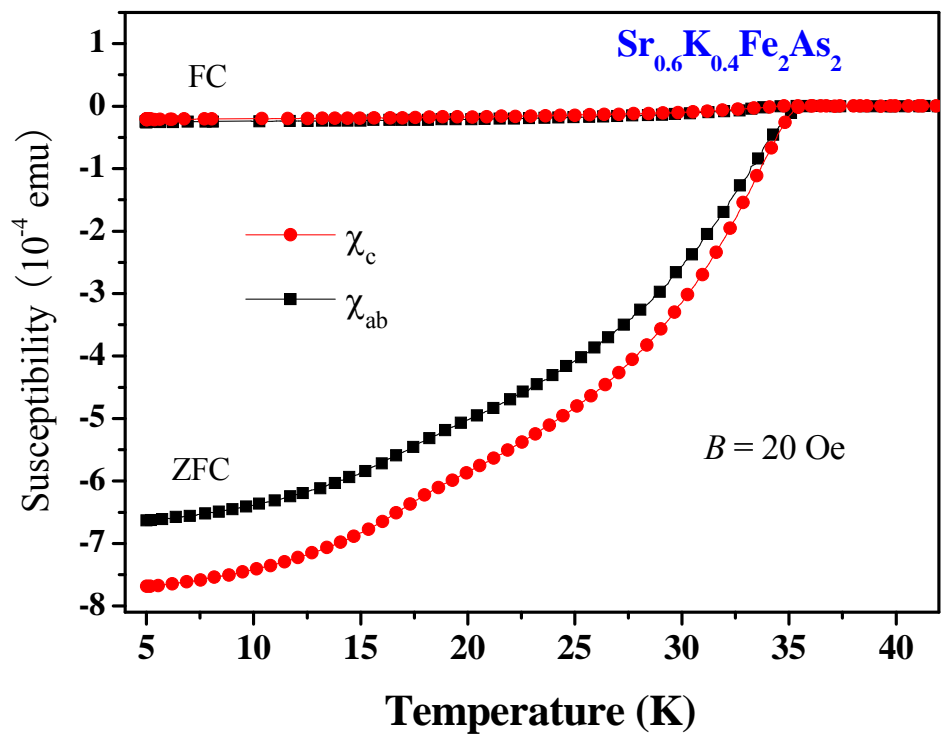

Fig.4 Gao et al.

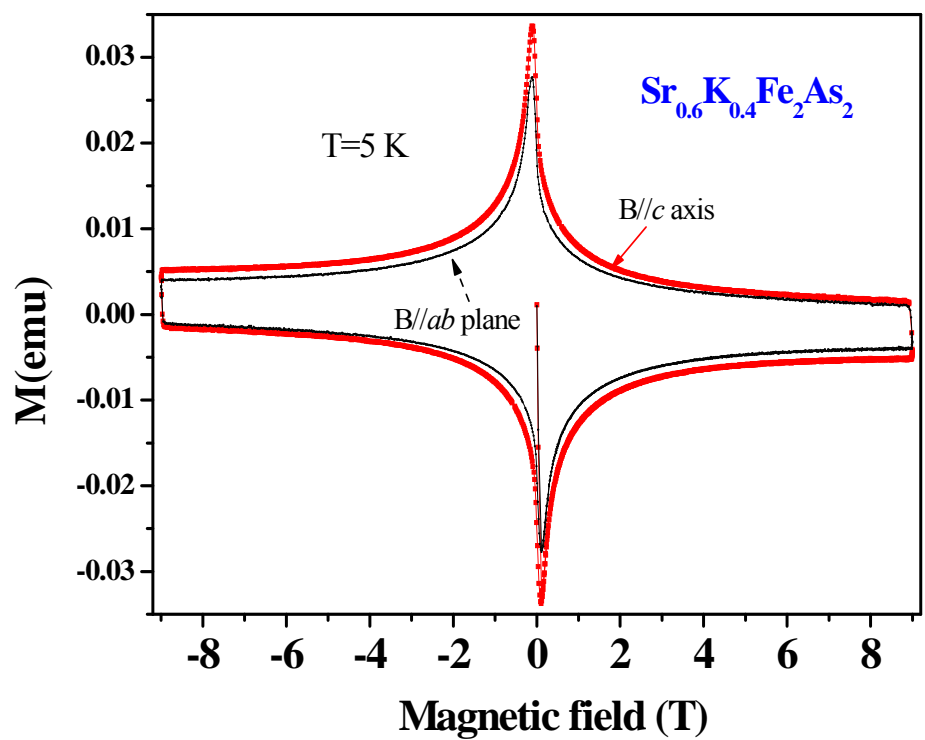

Fig.5 Gao et al.